\documentstyle[a4wide,epsf,11pt,titlepage]{article}

\newcounter{nref}
\newcommand{\bbib}{%
  \renewcommand{\refname}{\large\bf References}%
  \begin{thebibliography}{99}%
  \setcounter{enumiv}{\thenref}}
\newcommand{\ebib}{%
  \end{thebibliography}%
  \setcounter{nref}{\arabic{enumiv}}}
\newcommand{\head}[3]{%
  \setcounter{nref}{0}%
  \thispagestyle{empty}%
  \section*{\LARGE\bf #1}%
  \stepcounter{section}%
  \addcontentsline{toc}{section}{#1}%
  \large\itshape%
  #2\\\vspace{0.1pt}\\%
  #3%
  \normalsize\upshape%
  \bigskip}

\newcommand{\Msun}{{{\mathrm{M}_{\odot}}}}
\newcommand{\Sec}{{{\mathrm{s}}}}
\newcommand{\km}{{{\mathrm{km}}}}
\newcommand{\erg}{{{\mathrm{erg}}}}
\newcommand{\keV}{{{\mathrm{keV}}}}
\newcommand{\MeV}{{{\mathrm{MeV}}}}
\newcommand{\dex}{{{\mathrm{dex}}}}
\newcommand{\g}{{{\mathrm{g}}}}

\newcommand{\Ye}{{{Y_{\mathrm{e}}}}}
\newcommand{\bm}{{{\beta^-}}}
\newcommand{\bp}{{{\beta^+}}}
\newcommand{\ad}{{{\alpha}}}

\newcommand{\isofont}[1]{{\mathrm{#1}}}
\newcommand{\isomass}[1]{{{\isofont{^{#1}}}}}
\newcommand{\isocharge}[1]{{{\isofont{_{#1}}}}}
\newcommand{\isotope}[3]{{{\isocharge{#1}\isomass{#2}\isofont{#3}}}}
\newcommand{\I}[2]{{\isotope{}{#1}{#2}}}

\newcommand{\powersep}{{{\cdot}}}
\newcommand{\Ep}[1]{{{10^{#1}}}}
\newcommand{\E}[1]{{{\powersep\Ep{#1}}}}

\begin{document}

\head{Nucleosynthesis in Massive Stars \\ Including All Stable Isotopes}
     {A.~Heger$^1$, R.~D.~Hoffman$^2$, T.~Rauscher$^{1,3}$, 
     \&~S.~E.~Woosley$^1$}
     {$^1$ Astronomy Department, \\ University of California, Santa Cruz, CA~95064, U.S.A.\\
      $^2$ Nuclear Theory and Modeling Group, L-414, \\ Lawrence Livermore National Lab, Livermore, CA~94551-9900, U.S.A.\\
      $^3$ Departement f\"ur Physik und Astronomie, \\ Universit\"at Basel, CH-4056 Basel, Switzerland}

\subsection*{Abstract}

We present the first calculations to follow the evolution of all
stable isotopes (and their abundant radioactive progenitors) in a
finely zoned stellar model computed from the onset of central hydrogen
burning through explosion as a Type II supernova.  The calculations
were performed for a $15\,\Msun$ Pop I star using the most recently
available set of experimental and theoretical nuclear data, revised
opacity tables, and taking into account mass loss due to stellar
winds.  We find the approximately solar production of proton-rich
isotopes above a mass number of $A=120$ due to the $\gamma$-process.
We also find a weak s-process, which along with the $\gamma$-process
and explosive helium and carbon burning, produces nearly solar
abundances of almost all nuclei from $A=60$ to $85$.  A few
modifications of the abundances of heavy nuclei above mass $90$ by the
s-process are also noted and discussed. New weak rates lead to
significant alteration of the properties of the presupernova core.

\subsection{Introduction}

Stars above $\sim10\,\Msun$ are responsible for producing most of the
oxygen and heavier elements found in nature.  Numerous studies of such
stars and their detailed nucleosynthetic yields for various masses and
metallicities, have been carried out previously, e.g.,
\cite{alex:WW95,alex:TNH96}.  However, our knowledge of both the input
data and the physical processes affecting the evolution of these stars
has improved dramatically in recent years.  Updated opacity tables
\cite{alex:IR96} have become available along with more accurate
prescriptions for mass loss due to winds and new weak rates
\cite{alex:LM00} that affect the evolutionary stages after central
oxygen depletion. Perhaps most important for nucleosynthesis, new,
accurate reaction rates for all the relevant strong and
electromagnetic nuclear reactions above neon have been recently
calculated by Rauscher and Thielemann \cite{alex:RT00}.  Here we
present the first results for a $15\,\Msun$ supernova evolved with the
new physics. Additional masses and metallicities will be explored in
future papers. These future papers will also include rotationally
induced mixing processes \cite{alex:heg01}.

We also employ a nuclear reaction network of unprecedented size. The
nuclear reaction network used by \cite{alex:WW95} (WW95), large in its
day, was limited to 200 isotopes and extended only to to germanium
(see also Chieffi and Limongi in these proceedings).  Studies using
nuclear reaction networks of over 5000 isotopes have been carried out
for single zones or regions of stars, especially to obtain the
r-process, e.g., \cite{alex:CCT85,alex:fre99,alex:kra93}, but
``kilo-isotope'' studies of nucleosynthesis in complete stellar models
(typically of 1000 zones each) have been hitherto lacking.  We thus
also present the first calculation to determine, self-consistently,
the the complete synthesis of all stable isotopes in any model for a
massive star. However, because its thermodynamic properties continue
to be poorly determined (and for lack of space), we will ignore here
the nucleosynthesis that occurs in the neutrino wind, which may be the
principal site of the r-process \cite{alex:wwm94}.

\subsection{Input Physics}

Our calculations were performed using a modified version of the
stellar evolution code KEPLER \cite{alex:WZW78,alex:WW95} with the
following modifications:
\begin{itemize}
\item
updated neutrino loss rates \cite{alex:ito96}
\item
improved opacity tables (OPAL95) \cite{alex:IR96,alex:We96}
\item
mass loss due to stellar wind \cite{alex:NJ90}
\item
updated weak rates \cite{alex:LM00}
\item
updated strong and electromagnetic reaction rates \cite{alex:RT00}
\end{itemize}

As in WW95, nucleosynthesis was followed by co-processing the stellar
model throughout its evolution using an extended nuclear reaction
network.  From hydrogen ignition through central helium depletion a
617 isotope network was employed that included all elements up to
polonium, adequate to follow the s-process.  Just prior to central carbon
ignition, we switched to a 1482 isotope network (also including
astatine).  This new network incorporated more neutron-rich isotopes
to follow the high neutron fluxes in carbon (shell) burning.  Five to
ten isotopes were also added on the proton-rich side, in particular
for the heavy elements, to follow the $\gamma$-process
\cite{alex:WH78,alex:RAP90,alex:ray95}.  The nucleosynthesis during
the supernova explosion itself was followed in each zone using a 2439
isotope network that included additional proton-rich isotopes to
follow better the $\gamma$-process in the neon-oxygen core, and also
many additional neutron-rich isotopes to follow the n-process expected
during supernova shock front passage through the helium shell
\cite{alex:bla81}.

We implemented a new library of experimental and theoretical reaction
rates.  In particular, we used theoretical strong rates from
\cite{alex:RT00} (using input from the FRDM \cite{alex:moe95}), 
experimental neutron capture rates ($30\,\keV$
Maxwellian average) along the line of stability by \cite{alex:bao00},
and experimental
and theoretical rates for elements below neon as described in
\cite{alex:HWW00}. Experimental ($\alpha$,$\gamma$) rates were
implemented for $\I{70}{Ge}$
\cite{alex:fue96} and $\I{144}{Sm}$ \cite{alex:som98}. The derived
$\alpha$+$\I{70}{Ge}$ and $\alpha$+$\I{144}{Sm}$ potentials were also
utilized to recalculate the transfer reactions involving these potentials.

Experimental $\bm$, $\bp$, and $\ad$-decay rates were taken from
\cite{alex:NWC95}, experimental $\bm$ rates from \cite{alex:kra96} and
\cite{alex:Thi92}, and theoretical $\bm$ and $\bp$ rates from
\cite{alex:moe96}.  As a special case, we implemented the
temperature-dependent $\I{180}{Ta}$ decay as described in
\cite{alex:end99}.

The supernova explosion was simulated, as in \cite{alex:WW95}, by a
piston that first moved inward for $0.45\,\Sec$ down to a radius of
$500\,\km$ and then moved outward to a radius of $10\,000\,\km$ such
that a total kinetic energy of the ejecta at infinity of
$1.2\E{51}\,\erg$ resulted (for the $25\,\Msun$ stars we used a total
kinetic energy of $1.5\,\E{51}\,\erg$).  The final mass cut outside
the piston was determined by the mass that had settled on the piston
at $2.5\E4\,\Sec$ after core collapse.  Note that the amount of fallback
resulting from this prescription depends on both the initial location
of the piston used and the energy of the explosion.  In particular,
the yields of $\I{44}{Ti}$ and $\I{56}{Ni}$ are very sensitive to this
``final mass cut'' determined by the fall back.  Multidimensional
effects of the explosion are not considered here.  The temperature of
the $\mu$ and $\tau$ neutrinos emanating from the proto-neutron star
and causing the $\nu$-process nucleosynthesis \cite{alex:woo90} were
assumed to be $6\,\MeV$ in contrast to WW95 who assumed $8\,\MeV$.
However, we do not follow the $\nu$-process for isotopes with $Z$ or $N$
larger than $40$.

\begin{table}[b]
\caption{Properties of stellar models at the onset of core collapse
(\textit{first section}) and integrated stellar yields of some important radioactive nuclei
(\textit{second section}).%
\label{alex:models}}
\begin{tabular}{l@{}r|r@{}l@{}r@{}l@{}r@{}l|r@{}l@{}r@{}l@{}r@{}l}
\hline\hline
&& \multicolumn{6}{c|}{this work} & 
   \multicolumn{6}{c}{WW95 \cite{alex:WW95}} \\
\hline
initial mass  &($\Ep{34}\,\g)$& 3&   & 4&    & 5&    & 3&    & 4&    & 5&    \\
              & ($\Msun$)     &15&.08 &20&.11 &25&.14 &15&.08 &20&.11 &25&.14 \\
\hline  			     	    	                            
final mass    & ($\Msun$)     &12&.64 &14&.23 &13&.87 &15&.08 &20&.11 &25&.14 \\
He core       & ($\Msun$)     & 4&.16 & 6&.20 & 8&.19 & 4&.36 & 6&.67 & 9&.13 \\
C/O core      & ($\Msun$)     & 2&.82 & 4&.57 & 6&.38 & 2&.47 & 4&.37 & 6&.54 \\
Ne/O core     & ($\Msun$)     & 1&.87 & 2&.27 & 2&.77 & 1&.81 & 2&.44 & 2&.81 \\
Si core       & ($\Msun$)     & 1&.75 & 2&.07 & 2&.11 & 1&.77 & 2&.02 & 2&.06 \\
``Fe'' core   & ($\Msun$)     & 1&.55 & 1&.47 & 1&.74 & 1&.32 & 1&.74 & 1&.78 \\
delept. core  & ($\Msun$)     & 1&.29 & 1&.47 & 1&.59 & 1&.29 & 1&.74 & 1&.78 \\
central $\Ye$ &               & 0&.436& 0&.439& 0&.444& 0&.422& 0&.430& 0&.430\\ 
\hline				  	  	  	  	  	  
Pist. location& ($\Msun$)     & 1&.29 & 1&.47 & 1&.74 & 1&.29 & 1&.74 & 1&.78 \\
remnant mass  & ($\Msun$)     & 1&.72 & 1&.76 & 2&.31$^a$
                                                      & 1&.43 & 2&.06 & 2&.41$^b$ \\
\hline
\multicolumn{2}{c}{}&\multicolumn{12}{c}{radioactive yields} \\
\hline
$\I{26}{Al}$  & ($\Msun$)     & 4&.66$\E{-5}$ & 4&.89$\E{-5}$ & 1&.45$\E{-4}$ & 4&.30$\E{-5}$ & 3&.47$\E{-5}$ & 1&.27$\E{-4}$ \\
$\I{44}{Ti}$  & ($\Msun$)     & 1&.75$\E{-5}$ & 8&.89$\E{-6}$ & 2&.75$\E{-6}$ & 5&.68$\E{-5}$ & 1&.38$\E{-5}$ & 1&.95$\E{-6}$ \\
$\I{56}{Ni}$  & ($\Msun$)     & 9&.08$\E{-2}$ & 7&.67$\E{-2}$ & 5&.07$\E{-3}$ & 1&.15$\E{-1}$ & 8&.80$\E{-2}$ & 7&.26$\E{-5}$ \\
$\I{60}{Fe}$  & ($\Msun$)     & 7&.16$\E{-5}$ & 2&.84$\E{-5}$ & 1&.45$\E{-4}$ & 2&.66$\E{-5}$ & 1&.12$\E{-5}$ & 2&.10$\E{-5}$ \\
\hline
\end{tabular}
\\ $^a$ $1.5\E{51}\,\erg$ explosion
\\ $^b$ in \cite{alex:WW95} a remnant mass of $2.07\,\Msun$ is given in
error for the $1.2\E{51}\,\erg$ explosion.  This change decreases
significantly the yields of $\I{56}{Ni}$ and $\I{44}{Ti}$ (WW95:
$1.29\E{-1}\,\Msun$ and $3.04\E{-5}\,\Msun$, respectively).

\end{table}

\subsection{Results and Discussion}

\begin{figure}[t]
  \centerline{\epsffile{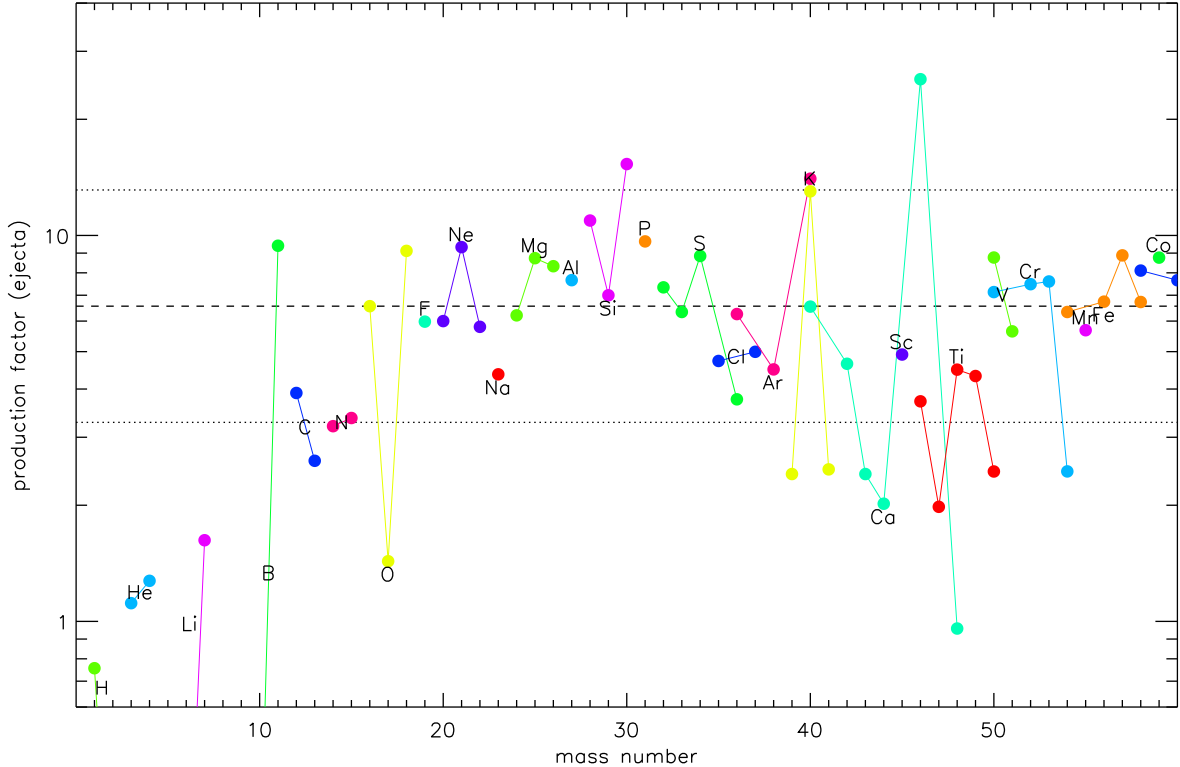}} \caption{Production
  factors of iron group and lighter nuclei in a $15\,\Msun$ star of
  solar metallicity.  Shown here are the average integrated abundances in the
  ejecta (including mass loss by stellar winds) relative to solar\cite{alex:GN93}
  (production factors).  The \textit{dashed line} indicates the
  production factor of $\I{16}O$ and the \textit{dotted lines} span a
  band of $\pm0.3\,\dex$. \label{alex:0-60}}
\end{figure}

\begin{figure}[t]
  \centerline{\epsffile{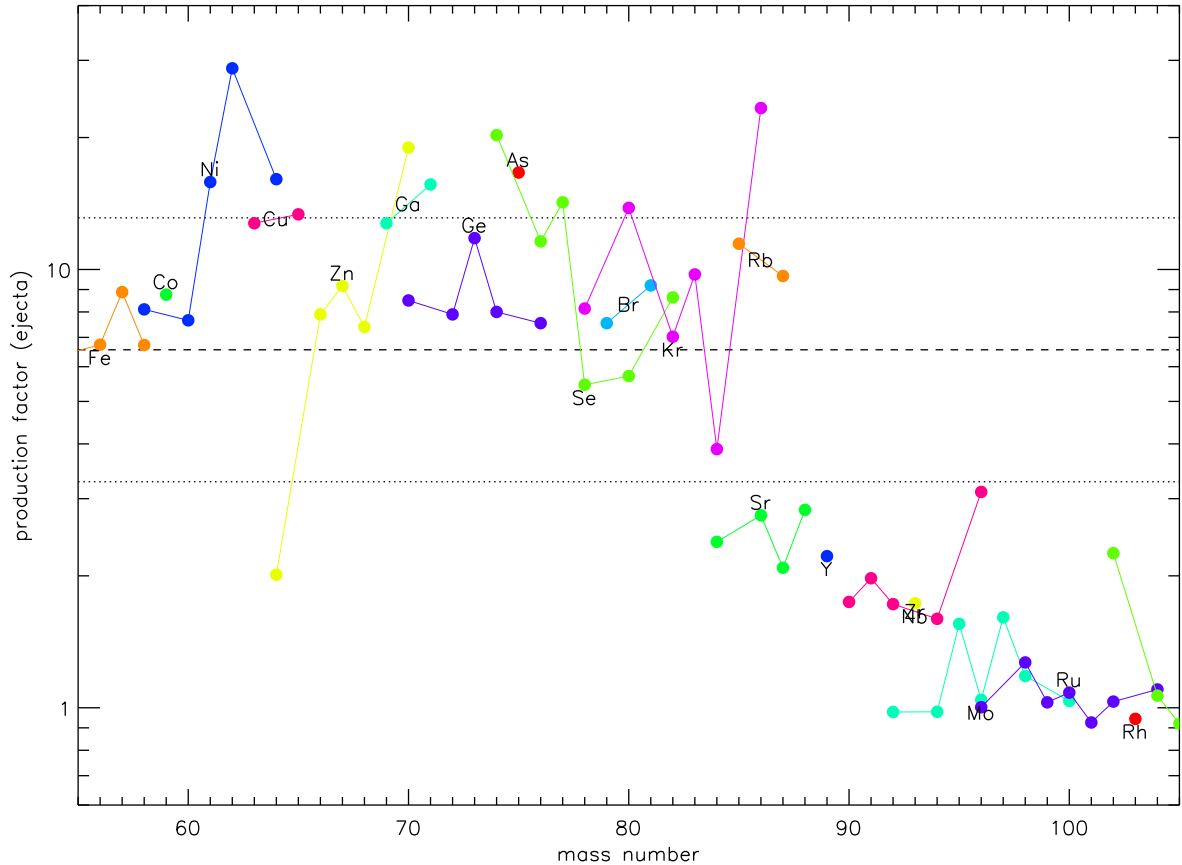}} \caption{Production
  factors of trans-iron group nuclei (all isotopes are on scale)
  \label{alex:55-105}}
\end{figure}

\begin{figure}[t]
  \centerline{\epsffile{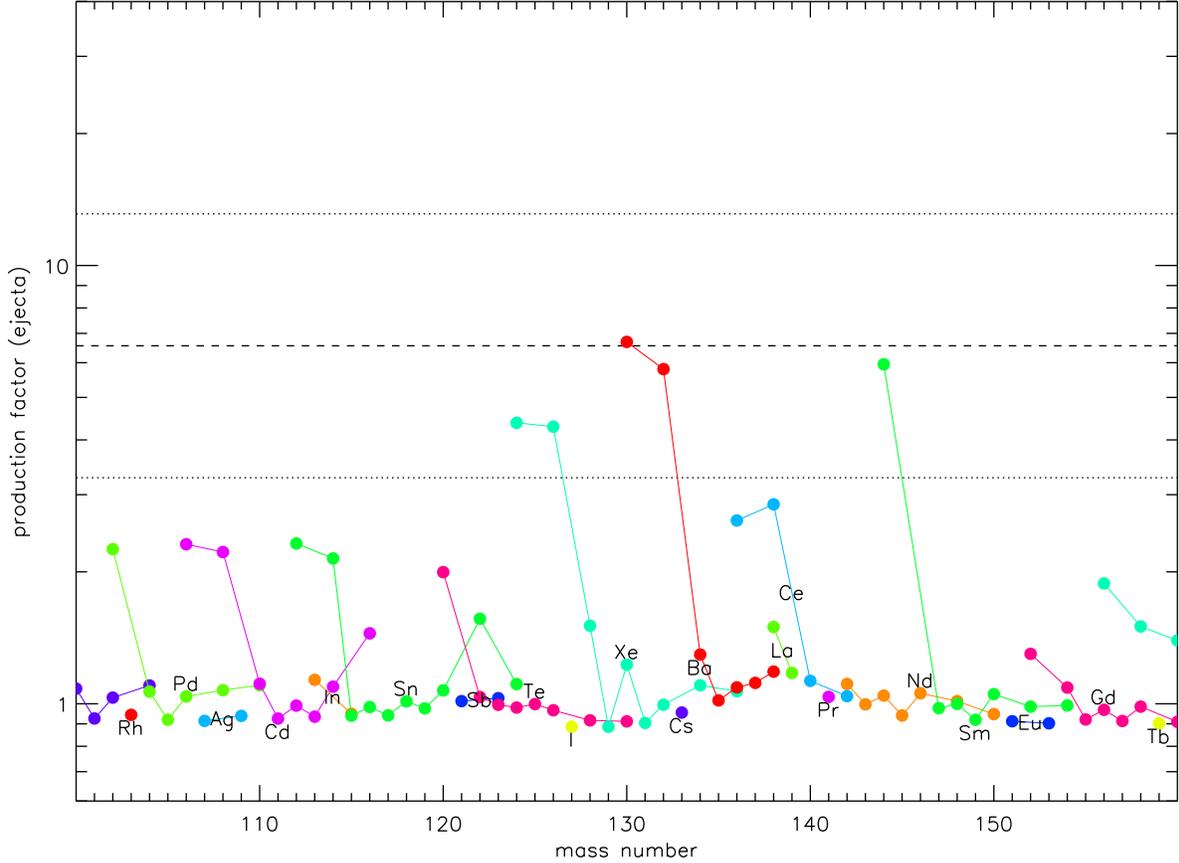}}
  \caption{Post iron group nuclei\label{alex:100-160}}
\end{figure}

\begin{figure}[t]
  \centerline{\epsffile{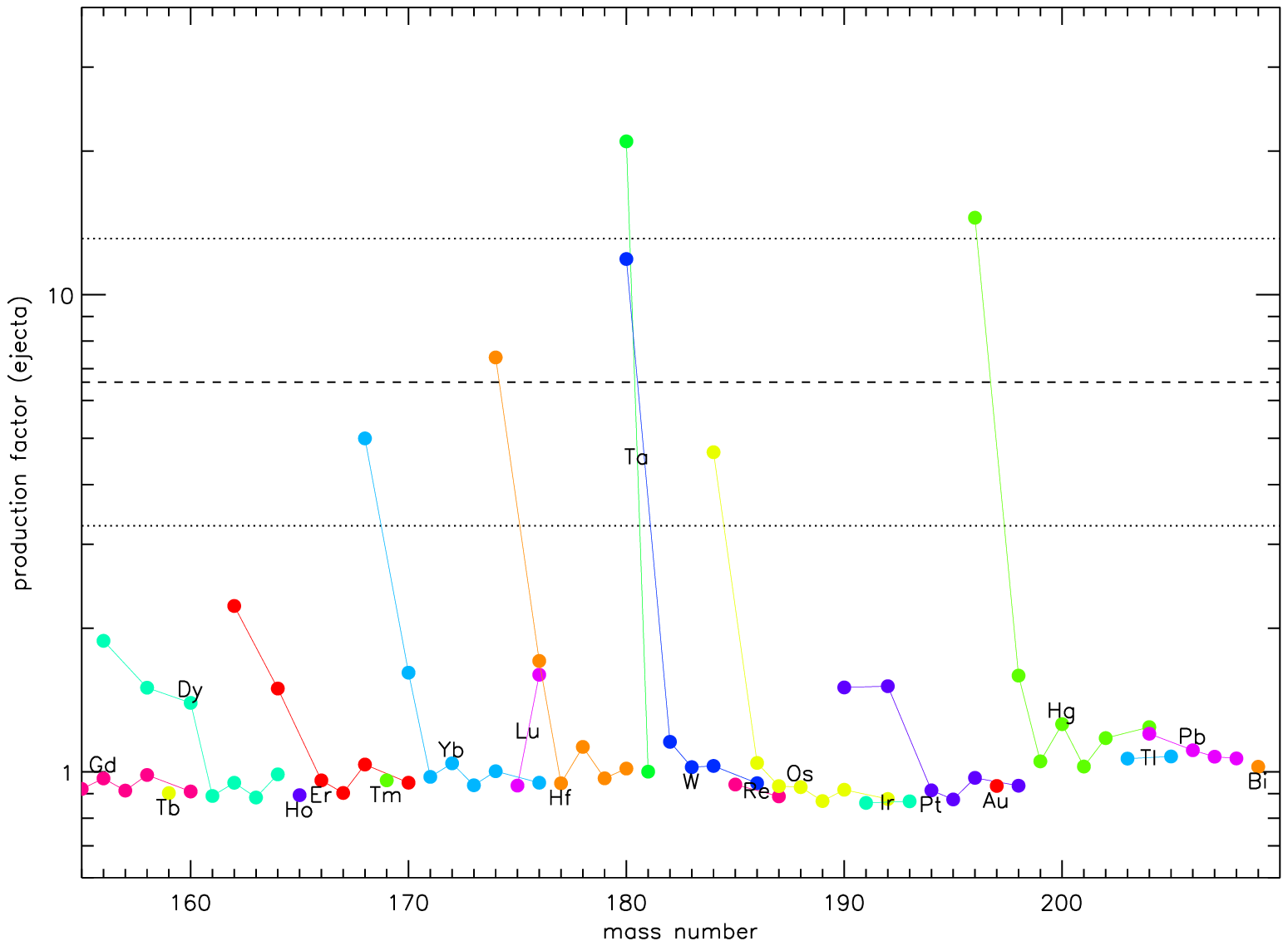}}
  \caption{Heavy nuclei\label{alex:155-210}}  
\end{figure}

\subsubsection{Stellar Structure}

Table~\ref{alex:models} gives the presupernova properties of our new
models and, for comparison, those of WW95.  The helium, carbon-oxygen,
and neon-oxygen cores were defined as the location where hydrogen,
helium, and carbon, respectively, first drop below a mass fraction of
$1\,\%$, from the stellar surface going inward.  The silicon core was
defined by where silicon becomes more abundant than oxygen and the
iron (``Fe'') core by where the sum of the mass fractions of
$\I{48}{Ca}$ and heavier nuclei first exceeds $50\,\%$.  The
deleptonized core was defined as the region where the number of
electrons per baryon, $\Ye$, first drops below 0.49.

The lower helium core masses in the new models
(Table~\ref{alex:models}) are due to both mass loss and the use of the
OPAL opacities.  In the $25\,\Msun$ case a model using OPAL opacities
but no mass loss resulted in a helium core of $8.69\,\Msun$.  As a
result of the reduced helium core size our new models typically have
lower helium-free and carbon-free cores.  Due to the interaction of
the different phases of shells burning, the sizes of the ``inner
cores'' do not always monotonically change with the size of the helium
core.  More details will be given elsewhere
\cite{alex:rau00,alex:heg01}.

Some important changes in the new models are due to the revised weak rates
\cite{alex:LM00}.  These rates become important during
core silicon burning and thereafter.  Typically, they lead to an
increase of the central $\Ye$ at the onset of core collapse by $2$ to
$3\,\%$ (Table~\ref{alex:models}), and this difference tends to
increase with increasing stellar mass \cite{alex:heg00}. Perhaps more
important for the explosion mechanism of core collapse supernovae is
an increase of the density in the mass range of $m=1.5\,\Msun$ to
$2\,\Msun$ by $30-50\,\%$ relative to the same models computed with
the previous set of weak rates \cite{alex:WW95}.  This may
significantly affect the dynamics of the core collapse.  For further
details see \cite{alex:heg00}.

\subsubsection{Nucleosynthetic Production Factors}

In Figs.~\ref{alex:0-60} through \ref{alex:155-210} we show the
production factors of all ejecta of the star after the explosion,
including all the mass lost due to stellar winds, relative to
solar \cite{alex:GN93} abundances.  We assume that all radioactive
nuclei have decayed to their stable products.  As a gauge we provide
the production factor of $\I{16}O$, the dominant ``metal'' produced in
massive stars (\textit{dashed line}), and a band of acceptable
agreement of $\pm0.3\,\dex$ relative to this values (\textit{dotted
lines}).

\subsubsection{Light Elements and the Iron Group}

The species $\I2H$, $\I3{He}$, lithium, beryllium and boron were
destroyed in the envelope of the star during central hydrogen burning.
However, substantial $\I7{Li}$ and $\I{11}B$ were recreated by the
$\nu$-process during the explosion \cite{alex:woo90}, as was $\I{19}F$
(Fig.~\ref{alex:0-60}).  $\I{17}O$ was significantly underproduced as
a result of the revised reaction rates for $\I{17}O(p,\alpha)\I{14}N$
and $\I{17}O(p,\gamma)\I{18}F$ \cite{alex:HWW00}.

The isotopes $\I{18}O$ through $\I{38}{Ar}$ are in good agreement with
solar abundance ratios.  $\I{40}{Ar}$ and $\I{40}K$ are both
significantly higher while other potassium isotopes are lower.  This
signature for the potassium was also found in other stellar models.
The under-abundance of $\I{44}{Ca}$ was caused by the low yield of
$\I{44}{Ti}$ (Table~\ref{alex:models}) which beta-decays to calcium.
The yield of this isotope strongly depends on the location of the
final mass cut, i.e., the amount of fall back, and might also be
affected by mixing processes during the supernova explosion.  The same
caveat also applies for other isotopes that mainly originate from
regions close to the neutron star, like $\I{56}{Ni}$ (see contribution
of Kifonidis in these proceedings).

\subsubsection{The s-Process}

The nuclei above the iron group up to about $A=90$
(Fig.~\ref{alex:55-105}) are produced as secondary isotopes by the
s-process starting from iron.  When considering galactic chemical
evolution these yields are to be combined with those of metal-poor
stars that contribute correspondingly less of these isotopes,
therefore a production factor of about twice that of $\I{16}O$ is in
good agreement with reproducing the solar abundance pattern.  Note
that the yields of these isotopes, by abundance, starting from iron
decreases about exponentially.  $\I{64}{Zn}$, which is underproduced as
shown in Fig.~\ref{alex:55-105}, may be a product of the neutrino wind
from the proto-neutron star \cite{alex:hof96}.  The possible
contributions due to this process are not included in the results
presented here.

The overabundance of the neutron-rich nickel isotopes,
$\I{61,62,64}{Ni}$, and other s-process products in the $A=60-90$ mass
range has been observed before \cite{alex:TWW95,alex:HWW00} and is
still not well understood. It is even greater in stars of 20 and
$25\,\Msun$. Perhaps the problem will be alleviated by a more complete
grid of supernovae of various metallicities and masses, perhaps the
stellar structure will be altered by still uncertain physics
(overshoot, $\I{12}C(\alpha,\gamma)\I{16}O$, rotation), or perhaps key
reaction rates responsible for neutron production or absorption will
change. For now, it remains problematic.

Above $A=100$ (Figs.~\ref{alex:100-160} and \ref{alex:155-210}) the
s-process had only minor effects in this $15\,\Msun$ star, though
there were important ``redistributions'' of some of the heavy
isotopes.  Most of the s-process above mass $90$ is believed to come
from AGB stars.

The ``cutoff'' towards lower values at a production factor of
$\sim0.9$ is due to the fact that most of the star does not become hot
enough to affect the abundances of these nuclei, or is even lost in
the wind.  The supernova ejecta containing regions depleted by the
s-process and other processes are then averaged with the dominating
contribution of unaffected matter.  In the $15\,\Msun$ star presented
here, this leads to the fact that $80\,\%$ of all \emph{ejecta},
including winds, did not experience the s-process.

\subsubsection{The $\gamma$-Process}

The production of the proton-rich nuclei results from
photo-disintegration of heavy nuclei during implosive and explosive
oxygen and neon burning ($\gamma$-process
\cite{alex:WH78,alex:RAP90,alex:ray95}).  Here we present the results
of the first calculations that follow the $\gamma$-processes through
the presupernova stages and the supernova explosion in the whole star.
Fully self-consistent, the $\gamma$-process here operates in stellar
regions that were exposed to previous episodes of s-processing.

Above $A=123$ to $A=150$ and between $A=172$ and $A=200$ the
proton-rich heavy isotopes are produced in solar abundance ratios
within about a factor of two relative to $\I{16}O$
(Figs.~\ref{alex:100-160} and \ref{alex:155-210}).  Below $A=123$ and
around $A=160$ the production of the proton-rich isotopes is down by
about a factor of three to four.  The total production of the
proton-rich isotopes increases for higher entropy in the oxygen shell,
i.e., with increasing mass of the helium core, as we have seen in our
$25\,\Msun$ star, but also depends on details of stellar structure and
the composition of the star at the time of core collapse.  Therefore
the contribution from more massive stars may well fill in the gaps of
the low production factors seen in the $15\,\Msun$ star.

The isotope $\I{180}{Ta}$, the rarest stable nuclear species
in the solar abundance
pattern, shows a remarkable overproduction (Table~\ref{alex:155-210})
in all of our models, despite our taking into account its destruction
by de-excitation into the short-lived ground state through thermal
excitation into an intermediate state \cite{alex:end99}.  This may
indicate that decay from other excited states could be important,
which are not accounted for here.  We cannot exclude, however, that
our treatment of $\I{180}{Ta}$ as a single species in the excited
state only may cause, at least in part, the overproduction found here.

\subsubsection{The r- and n-Process}

The base of the helium shell is suspected to be a possible site for
fast neutron capture processes as the supernova shock front passes
these layers, especially in the less massive core collapse supernovae.
Since current models of r-process sites have difficulties in
reproducing the r-process peak around $A=130$ when adjusted to fit the
heavier nuclei (see the contribution of Truran in these proceedings),
the base of the helium shell was considered as a possible environment
for producing these isotopes.

In our present models a distinct redistribution of nuclei around
$A=123$ was found at the base of the helium shell, but the resulting
yields were too small to constitute a significant contribution that
would be visible in Fig.~\ref{alex:100-160}.  We may speculate that less
massive core collapse supernovae might have a stronger contribution
though.  More details on the present calculations will be given in
\cite{alex:rau00}.

\subsection{Conclusions}

We have presented the first calculation to follow the complete
s-process through all phases of stellar evolution and the
$\gamma$-process in the whole star through the presupernova stage and
subsequent supernova explosion.  Below, we summarize the important
results for our $15\,\Msun$ star.  Note, however, that though this
mass is a numerically typical case of a Type II or Ib supernova, the
average nucleosynthetic yield of massive stars is the result of
populations of different stars each of which has its own peculiar
yields which must be combined to result in a solar-like abundance
pattern.  Some isotopes that are underproduced here may be strongly
overproduced in other massive stars while isotopes overproduced here
may be deficient in others.

The proton-rich heavy isotopes above $A=123$ can be well produced by
the $\gamma$-process occurring during implosive and explosive oxygen
and neon burning.  The proton-rich isotopes around $A=160$ and those
between $A=100$ and $A=123$, however, are underproduced by a factor of
$3$ to $4$ with respect to $\I{16}O$.  The isotope $\I{180}{Ta}$ shows a strong
overproduction by the $\gamma$-process.  This may indicate that decay
from excited states, of which we include only one, could be important.

A strong secondary s-process contribution appears between iron and a
mass number of $A=90$.  Above $A=100$ the s-process in our $15\,\Msun$
star is very weak, but it becomes notably stronger in stars with more
massive helium cores that perform helium burning at higher entropies.

The expected r- or n-process contribution due to the supernova shock
front running through the base of the helium shell does not show a
significant contribution in any of our preliminary model stars, not
even at $A=130$.  We observed some redistribution of isotopes at the
base of helium shell around $A=123$, but this did not show the
characteristics of a typical r-process nor was it important compared
to the total yield of the star.

The revisions of opacity tables and the introduction of mass loss
generally leads to smaller helium core sizes which tend to also
decrease the mass of the carbon-oxygen and the silicon core
(Table~\ref{alex:models}).  Note, however, that the absolute values of
these core masses depend on the uncertainties, in particular, of the
mixing processes in the stellar interior, such as semiconvection,
overshooting, and rotation.

The revision of the weak rates \cite{alex:LM00}, important after
central oxygen burning, leads to a $2-3\,\%$ higher electron fraction
per nucleon, $\Ye$, at the time of core collapse in the center of the
star (Table~\ref{alex:models}) and the ``deleptonized core'' tends to
comprise less mass \cite{alex:heg00}.  More important for the core
collapse supernova mechanism might be the $30-50\,\%$ higher densities
of the new models between the region of $m=1.5-2\,\Msun$
\cite{alex:heg00}, which may result in a correspondingly higher
ram-pressure of the infalling matter.

\subsection*{Acknowledgements}

We thank Karlheinz Langanke and Gabriel Mart{\'\i}nez-Pinedo for
discussion and supplying their theoretical weak reaction rates
\cite{alex:LM00} and are grateful to Frank Timmes for providing us
with his implementation of the neutrino loss rates of
\cite{alex:ito96} and the sparse matrix inverter we used for the large
network.  This research was supported, in part, by Prime Contract
No.~W-7405-ENG-48 between The Regents of the University of California
and the United States Department of Energy, the National Science
Foundation (AST 97-31569, INT-9726315), and the Alexander von
Humboldt-Stiftung (FLF-1065004).  T.R.~acknowledges support by a
PROFIL professorship from the Swiss National Science foundation (grant
2124-055832.98).



\bbib

\bibitem{alex:bao00} Z.Y.~Bao, H. Beer, F. K\"appeler, F. Voss, K. Wisshak,
and T. Rauscher, ADNDT \textbf{76} (2000) 1.

\bibitem{alex:end99} D.~Belic, et al., Phys. Rev. Lett. \textbf{83} (1999)
5242.

\bibitem{alex:bla81} J.B.~Blake, S.E.~Woosley, T.A.~Weaver, and
D.N.~Schramm, ApJ \textbf{248} (1981) 315.

\bibitem{alex:CCT85} J.J.~Cowan, A.G.W.~Cameron, and J.W.~Truran, ApJ
\textbf{294} (1985) 656.

\bibitem{alex:fre99} C. Freiburghaus, et al., ApJ \textbf{516} (1999) 381.

\bibitem{alex:fue96} Zs.~F\"ul\"op, et al., Z. Phys. A \textbf{355} (1996) 203.

\bibitem{alex:GN93} N.~Grevesse and A.~Noels, in Origin and Evolution
of the Elements, ed. N.~Prantzos, E.~Vangioni-Flam, M. Casse,
Cambridge, Cambridge Univ. Press (1993) p.~13.

\bibitem{alex:heg00} A.~Heger, S.E.~Woosley, G.~Mart{\'\i}nez-Pinedo,
and K.~Langanke, ApJ (2000) in prep.

\bibitem{alex:heg01} A.~Heger, S.E.~Woosley, R.D.~Hoffman, and
T.~Rauscher, ApJS (2001) in prep.

\bibitem{alex:hof96} R.D.~Hoffman, S.E.~Woosley, G.~Fuller, and
B.S.~Meyer, ApJ \textbf{460} (1996) 478.

\bibitem{alex:HWW00} R.D.~Hoffman, S.E.~Woosley, and T.A.~Weaver,
ApJ (2000) in prep.

\bibitem{alex:IR96}
C.A.~Iglesias and F.J.~Rogers, ApJ \textbf{464} (1996) 943.

\bibitem{alex:ito96} N.~Itoh, H.~Hayashi, A.~Nishikawa, and Y.~Kohyama,
ApJS \textbf{102} (1996) 411.

\bibitem{alex:kra96} K.L.~Kratz, et al., priv. com. (1996). 

\bibitem{alex:kra93}  K.L.~Kratz, et al., ApJ \textbf{402} (1993) 216.

\bibitem{alex:LM00} K.~Langanke and G.~Mart{\'\i}nez-Pinedo,
Nucl. Phys. \textbf{A673} (2000) 481.

\bibitem{alex:moe95} P.~M\"oller, J.R.~Nix, W.D.~Myers, and W.J.~Swiatecki,
ADNDT \textbf{59} (1995) 185.

\bibitem{alex:moe96} P.~M\"oller, J.R.~Nix, and K.L.~Kratz, ADNDT 
\textbf{66} (1997) 131. 

\bibitem{alex:NJ90}
H.~Nieuwenhuijzen and C.~de~Jager, A\&A \textbf{231} (1990) 134.

\bibitem{alex:NWC95} J.K. Tuli, et al.,
Nuclear Wallet Charts, 5$^{\mathrm{th}}$ edition, Brookhaven National
Laboratory, USA (1995).

\bibitem{alex:RT00} T.~Rauscher and F.-K.~Thielemann, ADNDT
\textbf{75} (No. 1+2) (2000) 1.

\bibitem{alex:rau00} T.~Rauscher, R.D.~Hoffman, A.~Heger, and
S.E.~Woosley, ApJ (2000) in prep.

\bibitem{alex:RAP90} M.~Rayet and M.~Arnould, and
N.~Prantzos, A\&A \textbf{227} (1990) 517.

\bibitem{alex:ray95} M.~Rayet, M.~Arnould, M.~Hashimoto, N.~Prantzos,
and K.~Nomoto, A\&A \textbf{298} (1995) 517.

\bibitem{alex:som98} E.~Somorjai, et al.,
A\&A \textbf{333} (1998) 1112.

\bibitem{alex:Thi92} F.-K.~Thielemann, priv. com. (1992). 

\bibitem{alex:TNH96} F.-K.~Thielemann, K.~Nomoto, and M.-A.~Hashimoto,
ApJ, \textbf{460} (1996) 408.  

\bibitem{alex:TWW95} F.X.~Timmes, S.E.~Woosley, T.A.~Weaver, and
S.E.~Woosley, G.~Fuller, and B.S.~Meyer, ApJS \textbf{98} (1995) 617.

\bibitem{alex:We96} A.~Weiss, priv. com. (1996).

\bibitem{alex:WH78} S.E.~Woosley and W.M.~Howard, ApJS \textbf{36}
(1978) 285.

\bibitem{alex:wwm94} Woosley, S. E., Wilson, J. R., Mathews, G., Meyer, B., 
and Hoffman, R. D., ApJ, \textbf{433} (1994) 229.

\bibitem{alex:WW95} S.E.~Woosley and T.A.~Weaver, ApJS \textbf{101}
(1995) 181.

\bibitem{alex:WZW78} T.A.~Weaver, G.B.~Zimmermann, and S.E.~Woosley, ApJ \textbf{225}
(1978) 1021.

\bibitem{alex:woo90} S.E.~Woosley, R.D.~Hoffman, D.~Hartmann, and 
W.~Haxton, ApJ \textbf{356} (1990) 272.

\ebib
\end{document}